

%
 \let\miguu=\footnote
 \def\footnote#1#2{{$\,$\parindent=9pt\baselineskip=13pt%
 \miguu{#1}{#2\vskip -7truept}}}
%
%


\def\implies{\Rightarrow}
\def\=>{\Rightarrow}
\def\==>{\Longrightarrow}
 
 \def\dal{\displaystyle{{\hbox to 0pt{$\sqcup$\hss}}\sqcap}}
 
%
\def\lto{\mathop
        {\hbox{${\lower3.8pt\hbox{$<$}}\atop{\raise0.2pt\hbox{$\sim$}}$}}}
\def\gto{\mathop
        {\hbox{${\lower3.8pt\hbox{$>$}}\atop{\raise0.2pt\hbox{$\sim$}}$}}}
%
 



\def\to{\rightarrow}		

\def\bar{\overline}		
\def\hat{\widehat}		


\def\ideq{\equiv}		


\def				
  \Complexes
   {{\rm C}\llap{\vrule height6.3pt width1pt depth-.4pt\phantom t}}


\def\tr{\rm tr}			



\def\interior #1 {  \buildrel\circ\over  #1}     





\def\BulletItem #1 {\item{$\bullet$}{#1}}


\def\AbstractBegins
{
 \singlespace                                        
 \bigskip\leftskip=1.5truecm\rightskip=1.5truecm     
 \centerline{\bf Abstract}
 \smallskip
 \noindent	
 } 
\def\AbstractEnds{\bigskip\leftskip=0truecm\rightskip=0truecm}

\def\ReferencesBegin
{
\singlespace					   
\vskip 0.5truein
\centerline           {\bf References}
\par\nobreak
\medskip
\noindent
\parindent=2pt
\parskip=4pt			        
 }

\def\section #1    {\bigskip\noindent{\bf  #1 }\par\nobreak\smallskip}

\def\subsection #1 {\medskip\noindent{\it [ #1 ]}\par\nobreak\smallskip}

\def\eprint#1{$\langle$e-print archive: #1$\rangle$}


\message{------------------------------------------------- }
\message{> > > > M a n u s c r i p t V e r s i o n : 9.3   }
\message{------------------------------------------------- }



\message{...assuming 8.5 x 11 inch paper}

\magnification=\magstep1	
\raggedbottom
\overfullrule=0pt 
\hsize=6.4 true in
 \hoffset=0.07 true in		
\vsize=8.8 true in

\voffset=0.125 true in         	

\parskip=9pt
\def\singlespace{\baselineskip=12pt}      
\def\sesquispace{\baselineskip=15.5pt}      



\def\rhohat{{\hat\rho\,}}
\def\Ehat{{\hat E}}
\def\Jhat{{\hat J}}
\def\Qhat{{\hat Q}}
\def\Hilb{{\cal H}}
\def\corr{\leftrightarrow}

\def\cita{\hangindent=15pt\hangafter=1}



\phantom{}
\vskip -1 true in
\medskip
\rightline{gr-qc/9705006}
\vskip 0.3 true in

\vfill

\bigskip
\bigskip

\sesquispace
\centerline{ {\bf THE STATISTICAL MECHANICS OF BLACK 
                HOLE THERMODYNAMICS}\footnote{*}%
{ To appear in the Proceedings of the Chandra Symposium: ``Black Holes and
  Relativistic Stars'', ed. R. M.~Wald, held December, 1996, Chicago,  
  (U. of Chicago Press). }}

\singlespace			        



\bigskip
\centerline {\it Rafael D. Sorkin}
\medskip
 

\smallskip
\centerline {\it Instituto de Ciencias Nucleares, 
                 UNAM, A. Postal 70-543,
                 D.F. 04510, Mexico}
 
 
\smallskip
\centerline { \it and }
\smallskip
\centerline {\it Department of Physics, 
                 Syracuse University, 
                 Syracuse, NY 13244-1130, U.S.A.}
\smallskip
\centerline {\it \qquad\qquad internet address: sorkin@suhep.syr.edu}

\AbstractBegins 
  Although we have convincing evidence that a black hole bears an entropy
  proportional to its surface (horizon) area, the ``statistical
  mechanical'' explanation of this entropy remains unknown.  Two basic
  questions in this connection are: what is the microscopic origin of the
  entropy, and why does the law of entropy increase continue to hold when
  the horizon entropy is included?  After a review of some of the
  difficulties in answering these questions, I propose an explanation of
  the law of entropy increase which comes near to a proof in the context of
  the ``semiclassical'' approximation, and which also provides a proof in
  full quantum gravity under the assumption that the latter fulfills
  certain natural expectations, like the existence of a conserved energy
  definable at infinity.  This explanation seems to require a fundamental
  spacetime discreteness in order for the entropy to be consistently
  finite, and I recall briefly some of the ideas for what the discreteness
  might be.  If such ideas are right, then our knowledge of the horizon
  entropy will allow us to ``count the atoms of spacetime''.  
\AbstractEnds

\sesquispace
\bigskip\medskip

When I first learned of the thermodynamics of black holes, and specifically
of the fact that a black hole possesses an entropy proportional to its
horizon area, my reaction (after thinking about it a while) was that this
was just as if the horizon were divided into small ``tiles'' of a fixed
size, with each tile carrying roughly one bit of information.  To see that
this would lead to the correct proportionality law, imagine a 0 or 1
engraved on each tile.  If there were $A$ tiles, then the number of
possible configurations would be
$N=2\times{}2\times{}2\cdots\times{}2=2^A$, and the corresponding entropy
would be $S = \log N = A\log{2}$ (taking Boltzmann's constant $k$ equal to
one). 
In order to get the numerical coefficient right, the tile size would have
to be around $10^{-65}$cm, that is, it would have to be of order unity in
units with $c=\hbar=8\pi{G}=1$.

Of course no one believes (or would admit to believing) that the black hole
horizon is painted with tiny 0's and 1's, but the suggestion remains that,
in some less artificial manner, a cutoff occurs at around the Planck scale,
and the microscopic degrees of freedom proper to the horizon carry about
one bit of information per horizon ``atom''.  In order for such an
explanation to work, it would have to provide convincing answers to two
principal questions: what degrees of freedom does this information capture
(to what $N$ is one really referring when one says $S=k\log{N}$), and why
does the total entropy still increase when black holes are involved
(why does the Second Law of Thermodynamics continue to hold)?

The explanation that would resolve these twin questions is what I mean by
the ``statistical mechanics behind black hole thermodynamics''.  We can
hope that in the process of arriving at such an explanation, we will learn
something important about the nature of spacetime on small scales, just as
the quest for the statistical mechanics of a box of gas taught us
something important about the nature of ordinary matter on atomic scales,
revealing the existence of atoms, their sizes, and something about their
structure and quantum nature.

Equally, we may hope that the investigation of these questions will shed
new light on statistical mechanics itself, and specifically on the meaning
of entropy.  For example, the experience so far has been that no derivation
of the Second Law can even get started without first applying some form of
coarse-graining to define the entropy, and the choice of coarse-graining
seems to introduce an unwelcome element of subjectivity into the
foundations of statistical mechanics.  But a black hole affords an obvious
objective way to coarse-grain, namely neglect whatever is inside the
horizon.  Or to take another example, it is the possibility of fluctuations
in the entropy that distinguishes the statistical mechanical picture from
the thermodynamical one, but in practice such fluctuations are ordinarily
too small to be observable.  With black holes, on the other hand, one can,
at least in principle, arrange for arbitrarily large fluctuations to occur
[1].

Before we address the statistical mechanics of black holes as such, let me
remind you very briefly of their thermodynamic properties in a little more
detail.  In a process like stellar collapse, a black hole is said to form
when a region of spacetime develops from which signals can no longer
escape.  Within the context of classical General Relativity the subsequent
occurrence of a singularity is then inevitable, but it is expected to form
inside the black hole, so that it is hidden from the view of distant
astronomers (see the discussion in [2]).

The black hole's {\it horizon} $H$ is by definition the three dimensional
surface separating the interior of the black hole from its exterior.
Formally, we have $H = \partial({\rm past}\, {\cal I^+})$, where
${\cal{I}^+}$ (called ``future null infinity'') is defined as the set of
ideal points at infinity at which outgoing light rays terminate.  Thus the
past of ${\cal I^+}$ consists of all events that can send light rays to
infinity; and $H$ is its boundary.  From this definition alone certain
mathematical facts follow, including the fact that $H$ is a continuous
surface (a $C^0$ manifold) which, although it may not be smooth (because of
caustics), nevertheless is null almost everywhere in the sense that it is a
union of null geodesics which never leave $H$ as they propagate into the
future.  (These ruling geodesics are photon world lines that hover on the
horizon, balanced precariously between escaping to infinity and being
pulled into the singularity.)  When we speak of the ``area of the
horizon'', we mean the area of the {\it cross-section} in which $H$
intersects some spacelike (or possibly null) hypersurface $\Sigma$ on which
we seek to evaluate the entropy:\footnote{*}
{For this definition of $A$ to make sense, it is necessary that
 $H\cap\Sigma$ not be too rough.  I don't know whether one can prove this
 for every possible black hole horizon $H$ and smooth surface
 $\Sigma$, but ``Geometric Measure Theory'' guarantees at least that, for
 any given $H$, there exists a dense set of $\Sigma$'s for which $A$ is
 well defined [3].}
$$
              A := Area (H\cap\Sigma) .
$$

Now an abundance of evidence indicates that there is associated with an
event horizon of this sort an entropy of
$$
    S_{BH} = { 2\pi A \over l^2 } ,   \eqno (1)
$$
where
$$
           l^2 = 8\pi G \hbar  \eqno (2)
$$
is the square of the ``rationalized Planck length''.  The best known
piece of evidence for this association is that the black hole radiates in
precisely the right manner to be in equilibrium with a surrounding gas of
thermal radiation at the temperature derived from (1) via the First
Law of Thermodynamics, $dM = T dS$ (the Hawking radiation).  But the
evidence goes well beyond that, ranging from theorems in classical General
Relativity that can be interpreted as the Zeroth through Third Laws of
Thermodynamics applied to a black hole in the $\hbar\to{}0$ limit, to
computations directly yielding the entropy (1) in the ``tree level''
approximation of path-integral quantum gravity.  

Many of these relationships carry over to other spacetime dimensions and
other gravity theories than standard General Relativity, as emphasized in
[4].  However, the evidence remains less complete and convincing
in these other cases.\footnote{*}
{In a sense it is disappointing that black hole thermodynamics appears to
be so robust: the less sensitive it is to the theoretical assumptions, the
less capable of guiding us to the correct underlying theory of quantum
gravity!}
In particular one still lacks an analog for these other cases of the result
that in standard General Relativity represents the $\hbar\to{}0$ limit of
the Second Law of Thermodynamics, namely the theorem that classically the
total horizon area necessarily increases (or remains constant) as the
hypersurface $\Sigma$ on which it is evaluated moves forward in time.
Because of the way that $\hbar$ enters into the expressions (2) and
(1), the black hole entropy goes to infinity in the classical limit,
and so tends to dominate all other entropies.  Therefore, this classical
law of area increase can be interpreted as precisely what remains of the
Second Law when $\hbar$ tends to zero.  (Unfortunately, the proof of area
increase remains incomplete even in standard General Relativity, because it
rests on the still unproven assumption of ``cosmic censorship''.)

Notice that the law of area increase, even though it is valid only in the
non-quantum limit, is a fully nonequilibrium result in the sense that no
requirement of stationarity is imposed on the black holes or the classical
matter with which they interact.  In the complementary limit of fully
quantum matter interacting with a near equilibrium black hole, there exist
several arguments and thought experiments suggesting the impossibility of
procuring a decrease in the total entropy (black hole plus entropy of
surroundings) by any process in which the black hole evolves
quasi-stationarily and remains essentially classical.  (Later, I will
propose a completely general proof of this impossibility.)  Taken together,
these 
results in special cases
strongly suggest that the Second Law of Thermodynamics,
$\Delta{S}\ge 0$, will continue to hold in general
if we attribute to each black hole
its
corresponding entropy (1) and take for the total entropy, the sum
of this horizon-area entropy with the entropy of whatever else may be
present {\it outside} the black holes:
$$
  S_{tot} = S_{BH} + S_{outside} \quad {\rm increases\ with\ time}. 
  \eqno(3)
$$

\noindent
{\it Remark}\phantom{A} The above discussion has defined a black hole in
causal 
terms, and correspondingly has taken the horizon to be what is more
precisely referred to as the {\it event horizon}.  A rather different
concept is that of {\it apparent horizon}, whose definition generalizes the
curvature properties of the Schwarzschild horizon, rather than its causal
properties.  In application to time-independent black holes, the two
concepts coincide, but away from equilibrium they differ.  It has been
proposed [5] to identify the area that enters into the
formula (1) as the area of the apparent horizon rather than the event
horizon.  The apparent horizon has the apparent advantage of being
definable ``quasi-locally'', whereas locating the event horizon requires in
principle a knowledge of the entire future of the spacetime.  On the other
hand, it is precisely this ``teleological'' attribute of the event horizon
that gives rise to the large entropy fluctuations alluded to above.
Moreover, the concept of event horizon seems more fundamental than that of
apparent horizon, and therefore more robust in relation to possible
discrete replacements for spacetime such as the causal set [6].  In
addition, the area of the apparent horizon will often jump discontinuously,
which is not how one might expect entropy to behave.  For these reasons, I
will stick with the identification, horizon = event horizon, but it seems
prudent to bear the other alternative in mind as well.

Now, how would we understand the nondecreasing character of $S_{tot}$ if a
black hole were an ordinary, nonrelativistic thermodynamic object?  For
example, suppose the black hole were really a warm brick or, say, a hot
ball of hafnium.  In that case (recalling the familiar plausibility
arguments, as nicely presented in [4]), we would identify
$N_{BH}$ (the subscripts stand for ``ball of hafnium'') as the number of
internal micro-states of the ball, and $S_{BH}=\log{N_{BH}}$ as its
logarithm.  If we went on to assume that the ball was {\it weakly coupled}
to its surroundings, then we could write (with $N_{out}$ being the number
of micro-states of the surroundings),
$$\eqalignno{
   N_{tot} &= N_{BH} \cdot N_{out}  &(4a)\cr
   S_{tot} &= S_{BH}  +    S_{out}  &(4b)\cr }
$$
(On the other hand, if the coupling was not weak, then the states of the
two subsystems would not be able to vary independently, and so the estimate
$N_{tot} = N_{BH} \cdot N_{out}$ would be inaccurate.)  Under the further
assumption that the dynamical evolution of the whole system was {\it
ergodic} (exploring all of the available state-space) and {\it unitary} (so
preserving state-space volume as measured by number of states), we would
conclude that the time spent in any region of state-space was proportional
to the number of states in that region:
$$
           dwell\ time\  \propto\ N_{tot} ,          \eqno(5)
$$
so that\footnote{*}%
{The plausibility argument here could be made stronger if we assumed a form
 of ``detailed balance'' for the effective Markov process operating on the
 macro-states, or better ``meso-states'' (cf. [7]).  Then it
 would follow immediately that the probability (per unit time, say) of a
 transition from state 1 to state 2, given that the system was currently in
 state 1, would be greater by $e^{S_2-S_1}$ than that of the reverse
 transition, given that the system was currently in state 2. \hfil\break
 \phantom{01234} It is interesting that such a Markovian model of the
 evolution appears easier to justify in a classical framework, where the
 system really is in a definite (albeit unknown) micro-state at any instant.
 In the quantum case, in contrast, there is no reason to believe that the
 state-vector normally lies within any particular eigensubspace of a
 macroscopic observable like the temperature distribution.  Thus, although
 the {\it definition} of entropy works better in the quantum case (because
 ``number of micro-states'' really makes sense there), the argument for its
 {\it increase} seems to proceed more happily in the classical setting.  To
 recover a Markovian picture in the quantum case as well, one might have
 recourse to a sum-over-histories interpretation of the formalism, or to
 the closely related picture 
 in which
 coarse-grained histories are obtained from
 sequences of projection operators as in [8].} 
the probability of a transition in which $N_{tot}$ increased would
be more likely than one in which it decreased, the discrepancy being
overwhelming for large entropy differences because
$\Delta{S}>>1\implies{}e^{\Delta{S}}>>>1$.

So what goes wrong with this reasoning if we try to apply it to a black
hole?  In the first place, it is at least peculiar that the number of black
hole states would be proportional to $e^{\rm Area}$ rather than $e^{\rm
Volume}$ as for other thermodynamic systems.  This peculiarity becomes more
troubling if we consider the example of the Oppenheimer-Snyder spacetime,
in which a Friedmann universe {\it of arbitrary size} is joined onto the
interior of a Schwarzschild black hole of arbitrary mass.  (See the remarks
in [9].)  The existence of such solutions means that the number of
possible {\it interior} states for a black hole is really infinite, which
is certainly not consistent with any formula like $S=\log{N}$.

A second set of problems concerns {\it ergodicity} and {\it internal
equilibrium}.  The course of events inside a collapsing star leads
classically to a singularity, and it is not at all obvious that this is
consistent with an ergodic exploration of all available states, including
for example the Oppenheimer-Snyder states just described.  But even if
quantum effects did restore ergodicity, there would remain a problem with
equilibrium.  Although I did not stress it above, an assumption of internal
equilibrium (or partial equilibrium) is needed in order to deduce the
entropy from the values of a few macroscopic variables.  For example,
knowing only the surface temperature of the ball of hafnium would not at
all allow you to deduce its entropy unless you added the assumption of
internal equilibrium;  if this assumption were mistaken,
 you might find the
apparent entropy suddenly starting to decrease, because the interior of the
ball was much colder than you had assumed.  In the same way, mere knowledge
of the external appearance of a black hole tells you little about its
interior, and since realistic black holes would seem to be far from
internal equilibrium, this is another reason to doubt whether the type of
state-counting utilized in (4) and (5) can 
carry over to black holes
with $N_{BH}$ interpreted as the number of interior states.

A related problem with such state-counting is that our earlier assumption
of {\it weak coupling} 
between the subsystem and its environment is doubly wrong
in the case of a black hole.  The coupling from outside to inside is not
weak but very strong, while the reverse coupling is not so much weak as
nonexistent!  Indeed this last observation points up the fact that
conditions in the interior should be irrelevant, almost by definition, to
what goes on outside.  And since the Second Law, as ordinarily formulated
for black holes, makes no reference to conditions inside, it seems
especially strange that it should have anything to do with counting
interior states.  (In contrast, the temperature distribution inside our
ball of hafnium can make a big difference in its interaction with the
outside world, as we have seen, and it is impossible to specify its entropy
without saying something about the internal conditions, if not explicitly,
then implicitly via the assumption of internal equilibrium.)

Finally there is the vexed question of {\it unitarity}.  Many people refer to
this in terms of an ``information puzzle'', the puzzle being that the facts
apparently belie their belief that there must be a well-defined unitary
S-matrix for the {\it exterior} region alone; but for our purposes here,
the existence of such an S-matrix is irrelevant, because all we are
interested in is the Second Law, and that certainly imposes no such
requirement.\footnote{*}
{On the contrary, a unitary evolution of the exterior region would be
inimical to the type of explanation of the Second Law I will propose
below.}
Rather, what the ordinary statistical mechanical reasoning requires is {\it
overall} unitarity (for exterior and interior regions together), and the
difficulty lies in combining this overall unitarity with the mode of
explanation that would locate the entropy of the black hole in the
multiplicity of its interior micro-states.  If the latter were correct,
then a contradiction with unitarity
would arise in the process of evaporation of a black
hole by Hawking radiation.  In fact, since the black hole loses entropy as
it shrinks (its surface area decreases), its number of internal states would
have to go down as well, and at some point there would no longer be enough
of them to support the correlations with the radiated particles which are
needed if the {\it overall} evolution is to remain unitary.

At least this conclusion is inevitable if we accept the semiclassical
description of the radiation as consisting of correlated pairs $A$ and
$\bar A$, the first of which goes off to infinity while the second falls
into the singularity.  In this approximation of quantum field theory on a
fixed, background black hole metric, the emitted particles $A$ taken alone
are described by a highly impure state of the exterior field, and the
unitarity of the overall quantum evolution is restored only when the
$A\corr\bar{A}$ correlations are taken into account.  (One
sometimes says that the particles $A$ are ``entangled'' with the particles
$\bar{A}$.)  

In order for these entanglement correlations to exist, however, it is
necessary that the interior of the black hole support a quantum state-space
of dimension at least $e^{S_{rad}}$, where $S_{rad}$ is the entropy of the
emitted thermal radiation.  If the number of internal states really
diminished as the black hole shrank then this would cease to be possible,
and so unitarity could be maintained only if the $A\corr\bar{A}$
correlations were transferred to ones of the form $A\corr{}B$ between {\it
emitted} particles.  Ultimately, if the black hole were allowed to
evaporate fully, then all of the $A\corr\bar{A}$ correlations would have to
be transferred to the outside; and thus unitary evolution of the whole
system, {\it together with}
the assumption that black hole entropy counts the
number of internal states, would require the external evolution {\it alone}
to be unitary, at least in the $S$-matrix sense.

Now, it is possible to imagine mechanisms by which the required $A\corr{}B$
correlations could arise, but it is much less easy to imagine how the
$A\corr\bar{A}$ correlations could disappear, since they arise in an
approximation which should remain good in full quantum gravity.  But if
they don't disappear then the quantum state would have to stay
``entangled'', which in turn would require the number of internal states to
remain much larger than allowed by the formula (1).  Hence one has
either to abandon the interpretation of black hole entropy in terms of
internal states or to grant that the overall quantum evolution of the
system, black hole + environment, is nonunitary (or both).

Despite the seeming inevitability of this conclusion, many workers still
hope to evade it, but in seeking to do so, they are driven to rather
desperate maneuvers, such as hypothesizing a new ``complementarity''
according to which the internal state-space would be of a different
dimensionality for observers outside the black hole than for observers
inside of it.  Such conceptual difficulties notwithstanding, the
attempt to imagine how the external region alone can possess a unitary
$S$-matrix (or perhaps even a dynamics which remains unitary during all
intermediate stages) has led to some interesting suggestions, including the
suggestion [10] that at very small distances the only
variables that remain are purely geometrical ones, with all other
distinctions (color, flavor, generation, etc.) being washed out.

By far the greatest effort toward providing a unitary statistical mechanics
for black hole thermodynamics has been exerted within the context of string
theory (which is not to say that strings are necessarily tied to a unitary
explanation, any more than any other approach is,
cf. [11]).  In fact most of the effort has been
directed to only one of the two questions I emphasized at the outset,
namely the question of what degrees of freedom one must count in order to
obtain the formula (1).  The earliest calculation of this sort that I
know of was performed by Steve Carlip [12] for a black hole in 2+1
dimensional gravity.  Although this calculation did not actually use a
stringy version of gravity, it used string technology to count certain
``gauge'' degrees of freedom defined on the horizon, obtaining (1)
with precisely the thermodynamically required coefficient.  This would tend
to agree with the suggestion that the entropy is localized at the horizon,
rather than inside the black hole.  More recently, some calculations in
string theory proper have also obtained equation (1) for zero
temperature black holes in certain higher dimensional theories of gravity
coupled to special combinations of gauge fields chosen to make
supersymmetric solutions possible.  These more recent calculations do not
actually work with black holes, rather they count certain ``membrane''
states in a flat-space limit and obtain a formula which can be interpreted
as the analytic continuation of (1) to that limit. (See [13]
for a review of this work.)

To the extent that the physical picture behind these stringy calculations
can be extrapolated from flat spacetime to genuine black hole geometries
(which assumes in particular that the ``phase-transition'' accompanying the
formation of the horizon would not interfere with analyticity), they
suggest that the degrees of freedom contributing to the entropy are
associated with certain types of membranes.  In contrast to Carlip's
calculation, they also could be taken to suggest that the relevant degrees
of freedom are those of the black hole as a whole (as opposed to being
localized at the horizon), but since there are no horizons in flat space,
such a suggestion would have to be very tentative at best.  Beyond this,
the stringy calculations do not (as far as I can see) shed much light on
the question that to me is most central about the entropy of black holes:
why is it finite at all?  We will see later that certain identifiable
contributions to the entropy that arguably must be present are infinite in
the absence of a short-distance cutoff at the horizon.  One might ask
whether these contributions are present in the string theory picture, and
if so, how the requisite cutoff arises.

Concerning the derivation of the Second Law and the objections raised
above, string theory apparently has little to say.  If, at the end of the
day, it is able to produce a unitary dynamics underlying quantum gravity,
and if it is able to count the states of a black hole in this framework
(and also explain why a weak coupling, equilibrium approximation is really
valid after all, etc.), then the derivation of the Second Law will be
reduced to the discussion one can find in any thoughtful textbook on
statistical mechanics.  However, we have seen that the demand for unitarity
in particular, seems to drive one to a new kind of inside--outside
complementarity that so far no one knows how to formulate.  For this reason
it seems impossible at this stage to address the question of the Second Law
within the confines of any unitary theory that identifies the black hole
entropy with the states of the black hole taken as a whole.  Let us return,
then, to our two main questions --- ``Why does S increase?'' and ``What does
S count?'' --- and consider them in the order just given, hoping that an answer
to the first will suggest an answer to the second.

\section{ Why does $S$ increase? }

What I want to describe now is an old proposal [14]
[15] that would derive the Second Law by appealing
directly to the property of black holes that makes them {\it different}
from all other objects, the fact that what goes on outside the horizon
takes place without any reference to what goes on inside.  Classically this
independence is exact, and quantum mechanically it can still be expected to
hold to an excellent approximation, despite some quantum blurring of the
horizon.  But if the exterior region really has a well-defined autonomous
dynamics, then it is particularly natural to ``coarse-grain away'' the
interior region and seek to identify the entropy that enters into the
Second Law with the entropy of some effective quantum density-operator
describing the exterior situation.\footnote{*}%
{This is not meant to imply that on philosophical grounds, we must neglect
the interior region because ``we'' cannot observe it.  Rather my attitude
would be that any coarse-graining at all is valid if it leads to a useful
definition of entropy.  The horizon is nevertheless special in this
connection because, as we will see, its special causal properties make it
possible to draw conclusions that could not be drawn with other types of
coarse-graining.}

More specifically, what we would like to do is evaluate the entropy on a
hypersurface $\Sigma$ like that discussed earlier, but with the difference
that our new $\Sigma$ terminates where it meets the horizon.  We would like
to associate an effective density-operator $\rhohat(\Sigma)$ with each such
hypersurface and to prove that
$$
            S(\Sigma'') \ge S(\Sigma') 
$$
whenever $\Sigma''$ lies wholly to the future of $\Sigma'$.  We will see,
in fact, that this approach to the question leads to two proofs of entropy
increase, a more fully worked out one which applies in the semiclassical
approximation
[15] (cf. the remarks in [16] and the related remarks in
[17] and [18]), 
and a more sketchy one which applies in full quantum
gravity [14] [15].
It also leads to the conclusions that the entropy is localized
(to the extent that entropy can ever be localized) at or just outside the
horizon [14] [19] [16], and
that $S$ owes its finiteness to a fundamental spacetime discreteness or
``atomicity'' [14] [19].

Let us take first the case of a quasi-classical, quasi-stationary black
hole, by which I mean that we ignore quantum fluctuations in $M$ and the
other black hole parameters and we assume that the spacetime geometry can
be well approximated at any stage by a strictly stationary metric.  [In
other words, we work in the framework of ``quantum field theory in curved
spacetime''.  Notice that the requirement of approximate stationarity
applies only to the metric; the matter-fields (among which we may include
gravitons) can be doing anything they like.]  In such a situation there
exists for the matter fields outside the black hole a unique thermal (or
``KMS'') state $\rhohat^0$, known as the ``Hartle-Hawking state'' and
expressible as 
$$
     \rhohat^0 \propto e^{ -\beta \Ehat } ,    \eqno(6)
$$
where $\Ehat$ represents the energy operator for the matter fields in the
external region and
$\beta$ is the reciprocal of the thermodynamic temperature of the black
hole.\footnote{*}
{This expression is inevitably not rigorous, if only because the theory of
(interacting) quantum fields is itself not rigorous.  A related comment is
that, in a nonstationary background, there would be difficulties with the
definition of $\Ehat$, even in the case of a free field, as Adam Helfer
pointed out to me after my lecture [20].}
(For simplicity, we may as well work with a nonrotating, charge-free black
hole.  Alternatively, one could just reinterpret $\Ehat$ as
$\Ehat-\omega\Jhat-\phi\Qhat$.)

Now let us limit our consideration of the entropy to the hypersurfaces
$\Sigma(t)$ of a foliation of the exterior region which is compatible with
the Killing vector field $\xi^a$ that expresses the stationarity of the
metric.  In other words, the $\Sigma(t)$ are the $t=constant$ surfaces of
some coordinate system for which the metric is explicitly time-independent.
(Notice, however, that it would {\it not} be appropriate to take $t$ to be
the Schwarzschild time coordinate, for example, since then the
cross-section $\Sigma(t)\cap H$ would not move forward along the horizon as
$t$ increased.)  To each hypersurface $\Sigma(t)$ there corresponds a
quantum Hilbert space in which the field operators residing on $\Sigma(t)$ are
represented (in particular the operator 
$\Ehat=\Ehat(t)$ of eq. (6)), and
in which the quantum state $\rhohat(t)$ can therefore be represented as a
density operator.  By identifying appropriately the hypersurfaces
$\Sigma(t)$ with one another (the natural identification here being that
induced by the Killing vector $\xi^a$), we identify their attached Hilbert
spaces, and the dynamical change of $\rhohat(t)$ with time becomes thereby a
motion of $\rhohat(t)$ within a single Hilbert space.  (Notice that this
evolution of $\rhohat$ is well defined because the boundary $H$ at which we
have truncated the hypersurfaces $\Sigma$ is an {\it event horizon}, across
which no information can propagate.  This property of autonomous evolution
would be lost if, say, $H$ were replaced by a timelike surface, or if we
tried to evolve $\rhohat(t)$ {\it backward} in time.)

Now the key feature of the state (6) for us is that it is
time-independent---as every equilibrium state must be by definition.  What
makes this so important is that given any well-defined ({\it but not
necessarily unitary}) evolution law for density-operators $\rhohat(t)$, and
given any state $\rhohat^0$ that is stationary with respect to this
evolution, we can find a function $f(\rhohat(t))$, defined for {\it
arbitrary} states $\rhohat(t)$, which is nondecreasing with time.  (More
precisely, this is true classically for an arbitrary Markov process, and
true quantum mechanically under a certain auxiliary technical assumption.)
In effect, every state ``wants to evolve toward the stationary one'', and
$f$ is a kind of Lyapunov functional measuring how close $\rhohat(t)$ has
come to $\rhohat^0$.  Now, when the stationary state has the Gibbsian form
(6), the quantity $f(\rhohat)$ turns out to be (up to an additive
constant)
$$
   S(\rhohat) - \beta < \Ehat > 
   \ = \  
   \tr \, \rhohat \, (\log\rhohat^{-1} - \beta \, \Ehat) ;
                                                   \eqno(7)
$$
that is, it turns out to be the free energy up to a factor of $-1/T$.

The proof of this little known result is so remarkably simple (at least
classically) that I cannot resist presenting it here, in the important
special case where $\beta=0$.  To begin with, note that the function
$f(x)\ideq{x}\ln{x}^{-1}$ is concave downward since $f''(x)=-1/x<0$ in the
range $0\le{}x\le{1}$.  Now consider a Markov process whose probability of
being in the $k^{th}$ state at some moment of time is $p_k$.  The quantity
(7), when $\beta=0$, is nothing but the entropy
$$
       S(p) = \sum\limits_k p_k \log p_k^{-1} = \sum\limits_k f(p_k) ,
$$
so what we have to prove is that $S(p)$ increases when the $p_k$
get replaced at some later time by $\sum_l T_{kl}p_l$, $T_{kl}$ being the
matrix of transition probabilities.  But we have
$$
\eqalign{
          S(p) & =   \sum\limits_k f(p_k)                  \cr
	       & \to \sum\limits_k f(\sum_l T_{kl}p_l)     \cr
	       & \ge \sum\limits_{kl} T_{kl} \, f(p_l)	   \cr
	       & =   \sum\limits_{l} f(p_l) ,              \cr }
$$
where to get the inequality, we used the concavity of $f$ together with the
fact that 
$$
          \sum\limits_l T_{kl} = 1
$$
because $T$ preserves (when $\beta=0$) the totally random state,
$p_k\ideq{1}$; and in the last step we used that
$$
          \sum\limits_k T_{kl} = 1
$$
(conservation of probability).  When the stationary state $p_k^0$ is not
uniform, the proof is almost as simple, only $S$ gets replaced by
$$
               \sum\limits_{k} p_k^0 \, f(p_k/p_k^0) ,
$$
which for a thermal $p^0_k=e^{-\beta E_k}$ is just
$$
  \sum p_k^0 {p_k\over p_k^0} \log {p_k^0  \over p_k} = 
  \sum p_k(-\beta E_k - \ln p_k) =
  S - \beta <E> ,
$$
which is indeed the classical form of (7).

Returning to the black hole situation, it is now simple to see that the
nondecreasing character of (7) entails that of the total entropy
in our semiclassical, quasistationary approximation.  To that end, consider
the small change of state that occurs as the hypersurface $\Sigma'$ moves
slightly forward in time to $\Sigma''$, and write now $S_{out}$ for the
entropy $S(\rhohat)$ of the exterior matter, in order to distinguish it
from the entropy $S_{BH}$ of the black hole itself.  For the latter we have
(from the ``First Law of black hole thermodynamics'') that
$$
       dS_{BH} = \beta \, dM ,                    \eqno(8)
$$
$M$ being the mass of the hole.  As part of the semiclassical
approximation, we also assume that the mass of the black hole adjusts
itself in response to the quantum mechanical {\it expectation value} of the
energy it exchanges with the matter fields.  Therefore (changing `$\Ehat$'
to `$\Ehat_{out}$' for notational consistency) we have\footnote{*}%
{One might worry that the two terms in this equation refer to two different
energies, $M$ to the total energy as defined at infinity, and $<\Ehat>$ to
an energy defined with respect to the stationary background geometry.  That
this is not really a problem can be seen, for example, by imagining an
infinitesimal mass $m$ falling into the black hole from infinity.  On one
hand, this augments the mass $M$ of the black hole by $m$; on the other
hand, by conservation of the conserved energy current $T^a_b\xi^b$ in the
stationary background, the exterior energy $E_{out}$ decreases by the same
amount $m$ as the mass passes through the horizon.  A more direct proof
that (classically) $dS_{BH}=-\beta dE_{out}$ may be extracted from the
derivation following eq. (13) in the first reference of [21].}
$$
       <\Ehat_{out}> + M = constant ,                \eqno(9)
$$
whence $dM = - d<\Ehat_{out}>$ and $dS_{BH}=-\beta\,d<\Ehat_{out}>$.
Putting this together with the fact that the
expression (7) must not decrease in the process (and observing
that $\beta$ in that expression represents a {\it fixed} parameter of the
stationary metric), we can write for the infinitesimal change
$\Sigma'\to\Sigma''$,
$$
\eqalignno{
              	& d (S_{out} - \beta   < \Ehat_{out} >) \ \ge \ 0  \cr
              	& d  S_{out} - \beta \, d<\Ehat_{out}>  \ \ge \ 0  \cr
		& d  S_{out} + dS_{BH}                  \ \ge \ 0  \cr 
		& d (S_{out} + S_{BH})       \ \ge \  0         &(10)\cr }
$$

Thus we have proved the Second Law in the semiclassical approximation, for
arbitrary processes in which the black hole geometry changes sufficiently
slowly.  This class of situations includes all that I know of for which
thought experiments have been done to check the Second Law (see
[4] for some of them); however it does not cover black holes
that are far from equilibrium, for example ones just formed by the
coalescence of two neutron stars and still in the process of settling down
to a stationary state.  Notice also that in this proof we are not {\it
deriving} the value of the horizon entropy, but only showing that the
Second Law holds {\it if} we use the value (1) provided by
thermodynamic arguments.

Finally, it should be added that the matter entropy $S(\rhohat)$ we have
been working with is actually infinite, due to the entanglement between
values of the quantum fields just inside and just outside the horizon.  In
a little while we will consider this near-horizon entanglement entropy as a
possible source of the black hole entropy itself, but for now it is just a
nuisance since the divergence in $S_{out}$ means that (7) diverges
as well.  Thus, making our proof rigorous would require showing that {\it
changes} in (7) are nevertheless well-defined and conform to the
temporal monotonicity we derived for that quantity.  This probably could be
done by introducing a high-frequency cutoff on the Hilbert space (using as
high a frequency as needed in any given situation) and showing that the
evolution of $\rhohat$ remained unaffected because the high-frequency modes
remained unexcited.\footnote{*}%
{Alternatively, perhaps one could show that the additive constant,
$-\log\,\tr\,{e^{-\beta H}}$, omitted from (7) canceled the
divergence 
in a well-defined sense.  In order to make the proof rigorous,
one would also have, for example, to specify an observable algebra for the
exterior fields and a representation of that algebra in which the operators
$\rhohat$ and $\Ehat$ were well-defined (which in particular might raise
the issue of boundary conditions near the horizon).  In addition, one would
have to prove that the autonomous evolution of $\rhohat$ was 
``completely positive'', this being a condition for the applicability of
the theorem 
that guarantees the nondecreasing nature of (7) in the quantum
case.
This probably could be done by expressing the autonomous evolution as a
unitary transformation followed by a tracing out of interior degrees of
freedom.  See [15] for references and some further discussion
of these points.}
	
To what extent can we expect to generalize these considerations to the case
of greatest interest, that of a fully dynamical, quantum black hole (or
holes)?  In one respect, the situation actually simplifies, because the
technical hypothesis of complete positivity is no longer needed.  The
complication, of course, is that the continued validity of some of our other
assumptions will depend on the structure of the quantum gravity theory in
which the proof is to be realized.  Without knowing that structure in
advance we can do no better than to list the features that would be needed
in order for the proof to go through.  In writing down this list, I will
assume that the hypersurfaces $\Sigma$ to which the entropy is being
referred can be specified by some generally covariant prescription that
continues to make sense in quantum gravity, at least in some sufficiently
great subset of the situations for which we would want to formulate a
Second Law.  (For example, we might take advantage of the fact that a
``box'' is needed in order for thermodynamic equilibrium to be possible,
and specify $\Sigma$ as the boundary of the future of a freely chosen
cross-section of the box, regarded as a timelike ``world tube'' of fixed
geometry. [More precisely, $\Sigma$ would be that part of the boundary lying
inside of the box and outside of any future horizons that were present.]
Such a $\Sigma$ would be ``achronal'', though not strictly spacelike, and
the temporal relationship of two $\Sigma$'s specified in this way would
follow directly from that of their cross-sections.)  With respect to some
such family of hypersurfaces the needed assumptions are these:
\BulletItem{
   there is defined a Hilbert space $\Hilb$, and for each surface
   $\Sigma$, an effective density-operator $\rhohat(\Sigma)$ acting in
   $\Hilb$ ,} 
\BulletItem{
            $\rhohat$ evolves autonomously in the sense that
            $\rhohat(\Sigma'')$ is determined by $\rhohat(\Sigma')$
            whenever $\Sigma''$ lies to the future of $\Sigma'$ , }
\BulletItem{
   there exists an operator $\Ehat$ defined at the boundary of the system
   (or in any case without direct reference to the region inside the black
   hole) and yielding the total conserved energy, }
\BulletItem{
	the operator $\rhohat_{E_0} = \theta(E_0-\Ehat)$ is preserved by
       the autonomous evolution, }
\BulletItem{
       $(\forall E_0) \, \dim(\Hilb_{\Ehat < E_0} ) < \infty$,
       where $\Hilb_{\Ehat < E_0}$ is the subspace of $\Hilb$ with total
       energy less than $E_0$. }

Depending on your expectations, hopes or fears for quantum gravity, you may
find different ones of these assumptions more or less plausible or
palatable.  For me all are at least plausible, albeit the entire framework
of Hilbert space and operator observables is unlikely to be reproduced in
an exact form in quantum gravity.  In any case, if we accept these
assumptions, then it follows easily that the quantity
$$
   S_{tot}\ideq\tr\rhohat\log{\rhohat}^{-1} \eqno(11)
$$
is a nondecreasing function of $\Sigma$.  (The underlying mathematical
theorem is proved in [22] and also in [15], where some
further discussion of the above assumptions may be found as well.)  This,
then, will establish the Second Law (3) if we can show in addition
that $S_{tot}$ is a sum of two terms, one associated with the horizon and
one with the exterior region.

\section{ Why is $S$ a sum?}

So why would $S_{tot}=\tr\rhohat\log{\rhohat}^{-1}$ take the form of the
sum $A/l^2+S_{surroundings}$ when $\rhohat$ describes the exterior state
including all fields, both gravitational and nongravitational?  For this
to occur, there would need to be a great many degrees of freedom just
outside or ``contained in'' the horizon (one might expect it to be
thickened due to quantum effects), making their own identifiable
contribution to $S_{tot}$.  And of course this contribution would have to
be proportional to the horizon area in order to agree with equation (1).
Three ideas that have been proposed for what the horizon degrees of freedom
might be are:
\item{(a)}{the geometrical shape of the horizon itself,}
\item{(b)}{the modes of quantum fields propagating just outside the horizon,}
\item{(c)}{the fundamental degrees of freedom of the substratum,}
\par\noindent
where by ``substratum'', I mean the underlying structure(s) of whatever
turns out to be the true theory of quantum gravity.  Of course these
possibilities are not necessarily exclusive of each other.  All three types
of contribution might be present, and they might also overlap significantly
(for example, substratum degrees of freedom might show up in an effective
description as geometrical variables describing the horizon shape).  In
concluding this article, I would like to say a few words about each of
these possibilities and how they bear on the question of spacetime
discreteness.

An attractive feature of the first proposal [14]
[23] is that it offers a geometrical explanation for the very
geometrical relationship (1).  Unfortunately, it is not easy to
estimate the magnitude of the quantum fluctuations in the horizon shape,
but there are indications  [21] that they become significant, not
at the Planck 
scale $l$ (as one might have expected), but at the much larger
scale $\lambda_0\sim(Ml^2)^{1/3}$.  Such fluctuations would
in effect spread the horizon into a shell of thickness $\lambda_0$
harboring a potentially unlimited source of entropy.  In fact, the density
of fluctuation modes diverges as their transverse wave-number goes to
infinity.  This means that, without any cutoff, the entropy residing in the
shape degrees of freedom would presumably be infinite.  On the other hand,
the very rapidity of the modes' growth rate also means that most of the
modes with wavelength greater than any specified lower bound
$\lambda_{min}$ have $\lambda\sim\lambda_{min}$.  In consequence, a result
like (1), (2) emerges automatically (at least in crudely
estimated order of magnitude) if one does assume a cutoff of magnitude
$\lambda_{min}\sim{}l$.
 
The second proposal leads to very similar conclusions.  Here, if we work in
the approximation of a fixed background geometry, and if we limit ourselves
to free quantum fields, then we can actually compute the entropy
$S_{out}=-\tr\rhohat\log\rhohat$ of the field, and we obtain
[14] [23] the result $S\sim{}A/\lambda_{min}^2$
where $\lambda_{min}$ is the cutoff.  In this case, we recognize the area
law explicitly, and we see again that we must choose $\lambda_{min}\sim{}l$
in order to recover an entropy of the correct order of magnitude.

Unlike for proposal (a), where the effect of shape fluctuations on the
density operator $\rhohat(\Sigma)$ is not completely clear, a contribution
of type (b) must necessarily be present in the entropy (11).  In
conjunction with the type (a) contribution, this leads to an argument for
the inevitability of spacetime discreteness: under the assumption of a true
continuum, the type (b) entropy could fail to be infinite only if it
provided its own cutoff by inducing, and thereby coupling to, horizon
fluctuations of type (a), but then these would take over and provide an
entropy that therefore would still be infinite.\footnote{*}%
{For a partly different interpretation see [24].\vskip-9pt }

Another facet of proposal (b) which should be mentioned here is the
apparent difficulty that, since each field makes its own contribution, the
black hole entropy would seem to depend on the number of species of
fundamental fields in nature.  However, this conclusion is inevitable only
if we ignore the coupling of field fluctuations to horizon shape and also
hold fixed the value of the cutoff $\lambda_{min}$.  But at fixed cutoff,
not only $S$ but also the Planck length (2) will be affected by the
addition of a new species, and it appears [25] that this dependence is
just what is needed to maintain the relationship (1) unchanged.

Of course the need for a cutoff bespeaks an underlying spacetime
discreteness, and so, to my mind, the most intriguing possibilities of type
(c) are those in which the substratum  has a discrete character.  In
such a theory one could hope to derive the entropy, not just by counting
discrete quantum states of physically continuous variables, but by counting
certain discrete physical elements themselves (compare the fact that the
entropy of a box of gas is, up to a logarithmic factor, just the number of
molecules it contains).  For example, in causal set theory, one might count
the number of causal links crossing the horizon (near $\Sigma$), or in
canonical quantum gravity in the loop representation, one might count the
number of loops cut by the horizon (within $\Sigma$).  In any such case,
the result would be something like the horizon area in units set by the
fundamental discreteness scale.  Thus, to reduce the evaluation of the
entropy (1) to a counting exercise of this sort would be to open up a
direct path to learning the value of the fundamental length.  And if that
were to occur, then the quest for the statistical mechanics of black hole
thermodynamics would certainly have led us to something of interest: it
would have led us to the atoms of spacetime itself.


\bigskip\noindent

This research was partly supported by NSF grant PHY-9600620.

\ReferencesBegin

\cita
[1]
  R.D.~Sorkin and D.~Sudarsky (in preparation).

\cita
[2] 
   R. Penrose, contribution to this volume.

\cita
[3] 
 Frank Morgan,
 {\it Geometric Measure Theory: A Beginner's Guide},
 second edition
 (Academic Press 1988), Section 4.11.  I thank G.J.~Galloway for this 
 information.

\cita
[4]
   R.M. Wald, contribution to this volume, and references therein;
``Entropy and Black Hole Thermodynamics'', 
  {\it Phys. Rev. D}{\bf  20}: 1271-1282 (1979).

\cita
[5]
  Sean A.~Hayward,
 ``General laws of black-hole dynamics'',
  {\it Phys. Rev. D} {\bf 49}:6467-6474 (1994).

\cita
[6]
 L.~Bombelli, J.~Lee, D.~Meyer and R.D.~Sorkin, 
 ``Spacetime as a Causal Set'', 
  {\it Phys. Rev. Lett.} {\bf 59}:521-524 (1987).

\cita
[7]
 R.D.~Sorkin, 
``Stochastic Evolution on a Manifold of States'',
          {\it Ann. Phys. (New York)} {\bf 168}: 119-147 (1986).

\cita
[8]
C.J.~Isham,
``Quantum Logic and the Histories Approach to Quantum Theory'',
  {\it J. Math. Phys.} {\bf 35}: 2157-2185 (1994)
  $<$e-print archive: gr-qc/9308006$>$.
  
\cita
[9] 
   R.D.~Sorkin, R.M.~Wald and Zhang Zh.-J.,
  ``Entropy of  Self-Gravitating Radiation'', 
   {\it Gen. Rel. Grav.} {\bf 13}:1127-1146 (1981).

\cita
[10] 
 G.~'t~Hooft,
``The Black Hole Interpretation of String Theory'',
 {\it Nuc. Phys. B}{\bf 335}: 138-154 (1990).

\cita
[11]
   S.W. Hawking, contribution to this volume.

\cita
[12]
S.~Carlip,
``Statistical Mechanics of the (2+1)-dimensional Black Hole'',
  {\it Phys. Rev.} D {\bf 51}: 632-637 (1995)
 $<$e-print archive: gr-qc/9409052$>$.

\cita
[13]
  G. Horowitz, contribution to this volume.

\cita
[14] 
  R.D.~Sorkin, 
 ``On the Entropy of the Vacuum Outside a Horizon'',
   in B. Bertotti, F. de Felice and A. Pascolini (eds.),
   {\it Tenth International Conference on General Relativity and Gravitation
   (held Padova, 4-9 July, 1983), Contributed Papers}, 
   vol. II, pp. 734-736
   (Roma, Consiglio Nazionale Delle Ricerche, 1983).

\cita
[15] 
 R.D.~Sorkin, 
``Toward an Explanation of Entropy Increase in the
    Presence of Quantum Black Holes'',
    {\it Phys. Rev. Lett.} {\bf 56}, 1885-1888 (1986).

\cita
[16]
 W.H.~Zurek and K.S.~Thorne,
 ``Statistical Mechanical Origin of the Entropy of a Rotating, Charged Black
   Hole'',
  {\it Phys. Rev. Lett.} {\bf 54}: 2171-2175 (1985).

\cita
[17]
 R.D.~Sorkin, 
``On the Meaning of the Canonical Ensemble'',
  {\it Int. J. Theor. Phys.} {\bf 18}:309-321 (1979).

\cita
[18]
 R.D.~Sorkin, 
``A Simplified Derivation of Stimulated Emission by Black Holes'', 
   {\it Classical and Quantum Gravity} {\bf 4}, L149-L155 (1987).

\cita
[19]
 G.~'t~Hooft,
 ``On the quantum structure of a black hole'', 
   {\it Nuclear Phys. B} {\bf 256}:727-745 (1985).

\cita
[20] 
A.D.~Helfer,					
``The Stress-energy Operator,
{\it Class. Quant. Grav.} {\bf  13}: L129-L134 (1996).

\cita
[21]
 R.D.~Sorkin,
``Two Topics concerning Black Holes: 
   Extremality of the Energy, Fractality of the Horizon'',
   in S.A.~Fulling (ed.), 
   {\it Proceedings of the Conference on Heat Kernel Techniques and Quantum
    Gravity, held Winnipeg, Canada, August, 1994}, pp. 387-407
   (Discourses in Mathematics and its Applications, \#4) 
   (University of Texas Press, 1995)
   $<$e-print archive: gr-qc/9508002$>$; and
``How Wrinkled is the Surface of a Black Hole?'',
  in David Wiltshire (ed.), 
  {\it Proceedings of the First Australasian Conference on General
       Relativity and Gravitation}, 
  held February 1996, Adelaide, Australia, pp. 163-174
  (University of Adelaide, 1996)
  $<$e-Print archive: gr-qc/9701056$>$;  %
A.~Casher, F.~Englert, N.~Itshaki, S.~Massar and R.~Parenti,
``Black Hole Horizon Fluctuations''
  $<$e-Print archive: hep-th/9606106$>$.

\cita
[22]
Woo Ching-Hung,
``Linear Stochastic Motions of Physical Systems'',
 Berkeley University Preprint, UCRL-10431 (1962).

\cita
[23]
 L.~Bombelli,  R.K.~Koul, Lee~J. and R.D.~Sorkin, 
``A Quantum Source of Entropy for Black Holes'', 
  {\it Phys. Rev. D} {\bf 34}, 373-383 (1986).

\cita
[24]
J.D.~Bekenstein,
 ``Do we understand black hole entropy?'',
   in {\it The Seventh Marcel Grossmann Meeting on Recent Developments in
    Theoretical and Experimental General Relativity, Gravitation and
    Relativistic Field Theories},
    proceedings of the MG7 meeting, held Stanford, July 24--30, 1994,
    edited by R.T.~Jantzen, G.~Mac~Keiser and R.~Ruffini
   (World Scientific 1996)
   $<$e-print archive: gr-qc/9409015$>$.

\cita
[25] 
L.~Susskind and J.~Uglum,	           
``Black hole entropy in canonical quantum gravity and superstring theory'',
  {\it Phys. Rev.} D {\bf 50}:2700-2711 (1994)
  $<$e-print archive: hep-th/9401070$>$;
T.~Jacobson,
``Black Hole Entropy and Induced Gravity'', 
\eprint{ gr-qc/9404039 }.

\end